\documentclass[twocolumn,aps,prl,showpacs,groupedaddress,amsfonts,amssymb,amsmath]{revtex4-1}
\usepackage{graphicx}
\usepackage{epstopdf}
\usepackage[colorlinks=true, letterpaper=true, pdfstartview=FitV, linkcolor=blue, citecolor=blue, urlcolor=blue]{hyperref}

\begin{document}

\draft

\title{Antichiral edge states and hinge states based on the Haldane model}
\author{Xiaoyu Cheng,$^{1,2}$ Jun Chen,$^{3,2,*}$ Lei Zhang,$^{1,2,\dagger}$ Liantuan Xiao,$^{1,2}$ Suotang Jia$^{1,2}$}
\address{$^1$State Key Laboratory of Quantum Optics and Quantum Optics Devices, Institute of Laser Spectroscopy, Shanxi University, Taiyuan 030006, China\\
$^2$Collaborative Innovation Center of Extreme Optics, Shanxi University, Taiyuan 030006, China\\
$^3$State Key Laboratory of Quantum Optics and Quantum Optics Devices, Institute of Theoretical Physics, Shanxi University, Taiyuan 030006, China}

\begin{abstract}
Different from the chiral edge states, antichiral edge states propagating in the same direction on the opposite edges are theoretically proposed based on the modified Haldane model, which is recently experimentally realized in photonic crystal and electric lattice systems. Here, we instead present that the antichiral edge states in the two-dimensional system can also be achieved based on the original Haldane model by combining two subsystems with the opposite chirality. Most  importantly, by stacking these two-dimensional systems into three-dimension, it is found that the copropagating antichiral hinge states localized on the two opposite diagonal hinge cases of the system can be implemented. Interestingly, the location of antichiral hinge states can be tuned via hopping parameters along the third dimension. By investigating the local Chern number/layer Chern number and transmission against random disorders, we confirm that the proposed antichiral edge states and hinge states are topologically protected and robust against disorders. Our proposed model systems are expected to be realized in photonic crystal and electric lattice systems.
\end{abstract}
\maketitle
Recently, there has been emerging interests in the study of high-order topological insulators, which are proposed and have attracted widespread research attention in the fields of condensed matter physics\cite{langbehn2017reflection,schindler2018higher,luo2019higher,tanaka2020theory,khalaf2018higher,ezawa2018strong,kunst2018lattice,schindler2018higher1,song2017d,ezawa2018higher,wang2021transport,ren2020engineering,li2021higher,wang2020disorder,wang2021robustness,liu2019helical}, optics\cite{benalcazar2017quantized,chen2019direct,chen2020Antichiral,Xie2018Second,xie2019visualization,noh2018topological}, acoustics\cite{zhang2019dimensional,zhang2019second,ni2019observation,serra2018observation,xue2019acoustic,yang2021hybrid}, mechanics\cite{susstrunk2015observation}, etc. In the conventional topological insulators, the topologically protected boundary states appear with one dimension lower than that of the corresponding bulk system\cite{hasan2010colloquium,qi2011topological}. In contrast, the $n$-order high-order topological insulator (HOTI) in $d$ dimension can support $(d-n)$-dimensional topological boundary states, which can be two or more dimensions lower than that of the corresponding bulk system\cite{langbehn2017reflection,schindler2018higher,luo2019higher,tanaka2020theory,khalaf2018higher,ezawa2018strong,kunst2018lattice,schindler2018higher1,song2017d,ezawa2018higher,wang2021transport,ren2020engineering,li2021higher,wang2020disorder,wang2021robustness,liu2019helical,benalcazar2017quantized,chen2019direct,chen2020Antichiral,Xie2018Second,xie2019visualization,noh2018topological,zhang2019dimensional,zhang2019second,ni2019observation,serra2018observation,xue2019acoustic,yang2021hybrid,susstrunk2015observation}. Up to now, the existence of zero energy corner states in the second-order topological insulators are already experimentally realized in many different systems, for example, electrical circuits\cite{imhof2018topolectrical,shang2020second,serra2019observation}, photonics\cite{Xie2018Second,noh2018topological}, acoustics\cite{serra2018observation,yang2021hybrid}, etc.

Lately, the concept of antichiral edge states in the two-dimensional system is theoretically proposed based on a modified Haldane model\cite{colomes2018antichiral}, which alters the next-nearest-neighbor hopping parameters in both A and B sublattices. According to the chirality of the conventional chiral edge states, the edge states on the opposite edges must propagate in the opposite direction. However, the antichiral edge states on the opposite edges can propagate in the same direction instead. This interesting phenomenon has also been theoretically investigated in many other physical systems, for example, an exciton-polariton honeycomb lattice with strip geometry\cite{mandal2019antichiral}, a Heisenberg ferromagnet on the honeycomb lattice\cite{bhowmick2020antichiral}, and a graphene nanoribbon with zigzag edges under a uniform uniaxial strain\cite{mannai2020strain}, etc\cite{denner2020antichiral,yu2021antichiral,mizoguchi2021bulk,de2021odd}. More importantly, the existence of antichiral edge states based on this modified Haldane model has been experimentally demonstrated in gyromagnetic photonic crystal system\cite{zhou2020observation} and classical circuit lattice\cite{yang2020observation}. It is found that the robust propagation of antichiral edge states topologically protected by the winding number depends on the corner shape in the open boundary condition\cite{zhou2020observation}. Then the natural question arises: Is it possible to realize the antichiral edge states in the two-dimensional topological insulator and antichiral hinge states in three-dimensional high-order topological insulator based on the original Haldane model?

In this Letter, we answer this question in an affirmative way by realizing the antichiral edge states with the original Haldane model by combining two subsystems with opposite chirality. Conceptually, the realization of antichiral edge states here is totally different from that based on the modified Haldane model, i.e., the translational symmetry is broken due to the construction of two different subsystems. The proposed antichiral edge states will appear as midgap states in the gapped bulk states. By calculating the local Chern number, we find that the antichiral edge states are topologically protected. Furthermore, the disorder averaged transmission $\langle T \rangle$ is found to be quantized over a wide range of random disorder strength, which verify the robustness of antichiral edge states. Most importantly, the antichiral hinge states can be realized by stacking the two-dimensional system to the third dimension, which forms a antiferromagnets (AFMs) system characterizing by the slab Chern number. The robustness of antichiral hinge states is further confirmed by calculating the disorder averaged transmission. Moreover, the location of antichiral hinge states can be further tuned by varying the hopping parameters between two-dimensional layered systems.

\begin{figure}
\centering
\includegraphics[scale=0.45]{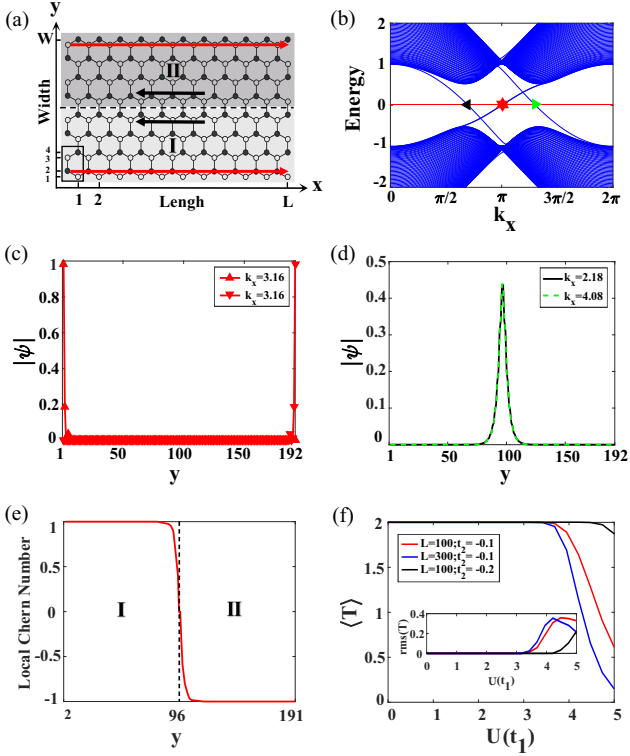}
\caption{(a) Schematic plot of zigzag nanoribbon system with width $W$ and length $L$ consisting of two regions I (light gray) and II (dark gray). Hollow circles and solid-filled circles represent A and B sublattice sites of the honeycomb lattice, respectively. The black box at the bottom left corner represents a unit cell in the system. (b) Band structure of the corresponding zigzag nanoribbon system in (a) with width $W = 192$. (c) The absolute value of wave functions with energy $E=0.01$ along transverse $y$ direction for two antichiral modes denoted as red up-pointing and down-pointing triangles in (b). (d) The absolute value of wave functions with energy $E=0.01$ for two counterpropagating modes denoted as black left-pointing and green right-pointing triangles in (b). (e) Local Chern number versus transverse $y$ direction. (f) The disorder averaged transmission coefficient $\langle T \rangle$ and its fluctuation $\textrm{rms}(T)$ versus disorder strength $U$ with system size $W=192,L=100,300$. Each data point on the figure is averaged over $2000$ disorder configurations. Here, $t_{1}=1$ is taken as the unit of energy, $t_{2} = -0.1$.}
\label{Fig1}
\end{figure}

First, we propose a model system (honeycomb lattice along the zigzag direction) consisting of two regions (I and II) along the $y$ direction as shown in Fig. \ref{Fig1}(a) to realize the antichiral edge states. The proposed Hamiltonian can be expressed as,
\begin{equation}
H_{2D} = H_{I} + H_{II} + H_{I,II},\label{Ham2D}
\end{equation}
where individual I/II region's Hamiltonian $H_{I/II}$ and coupling Hamiltonian $H_{I,II}$ between two regions are given by
\begin{equation}
\begin{split}
H_{I/II} &= t_{1}\sum_{\langle ij \rangle}c_{i}^\dagger c_{j} + t_{2}\sum_{\langle \langle ij\rangle\rangle }e^{-i\nu_{ij}\phi_{I/II}}c_{i}^\dagger c_{j},\\
H_{I,II} & = t_{1}\sum_{\langle i \in I, j\in II \rangle}c_{i}^\dagger c_{j},
\end{split}\label{Ham}
\end{equation}
where $c_{i}^\dagger$ and $c_{i}$ are the creation and annihilation operators at site $i$, $i$ and $j$ run over all sites in the system, $t_{1}$ is the nearest-neighbor hopping strength, $t_{2}$ describes the next-nearest-neighbor hopping, and $\nu_{ij} = \pm1 $ corresponds with counterclockwise/clockwise hopping of A/B site in the hexagonal lattice, $\phi_I=-\phi_{II}=\pi/2$. In order to account for the random disorder effect, we consider the static Anderson-type disorder to the on-site energy with a uniform distribution between $[-U/2,U/2]$.

Note that the Hamiltonian $H_{I/II}$ in Eq. (\ref{Ham}) in each region is based on the Haldane model, which can generate the quantum anomalous Hall effect (QAHE) when $|\phi_{I/II}|=\pi/2$\cite{haldane1988model}. To verify the topological property of the studied finite system, we calculate the local Chern number $\nu({\bf r})$ by the antisymmetric product of the projection operators\cite{mitchell2018amorphous,kitaev2006anyons,jiang2019experimental}. The size of finite-sample is 50 $\times$ 192 sites and the numerical results of local Chern number $\nu$ versus the transverse direction are given in Fig. \ref{Fig1}(e). It is found that the local Chern number (detailed calculations are provided in Supplemental Material\cite{suppsee}) $\nu$ changes from 1 in region I to -1 in region II along the transverse $y$ direction, which indicates that there must exist topologically protected edge states in regions I and II. Due to the opposite sign of local Chern number in two regions, the corresponding chirality of the edge states in regions I and II should also be reversed to each other. Physically, the edge states in the upper and lower boundary of the system have the same propagating direction while the counterpropagating states should be located in the centrally transverse direction as schematically shown in Fig. \ref{Fig1}(a). Thus, the antichiral edge states can be achieved in this combined two region system. By first investigating the band structure of a system with $W=192$ shown in Fig. \ref{Fig1}(b), we find that there are four states in the band gap. Through velocity analysis, it is easily found that two of them are propagating along positive $x$ direction and degenerate, while the other two states have counterpropagating direction. Furthermore, the absolute value of four wave functions with $E=0.01$ along the transverse direction are presented in Figs. \ref{Fig1}(c) and \ref{Fig1}(d). Indeed, two degenerate states are located in the upper and lower boundary, which indicate that they are antichiral edge states. The other two counterpropagating states are mainly distributed in the center of the transverse $y$ direction.

In addition, we simulate the quantum transport properties of antichiral edge states by contacting the system with two semi-infinite leads. Here the random disorders are in the presence of the central region. By using nonequilibrium Green's function method\cite{datta1997electronic,xing2011topological,zhang2014universal,khomyakov2005conductance,wang2009relation,sorensen2009efficient,zhang2012first}, the transmission can be calculated and averaged over disorder configurations (detailed information on how to calculate transmission $T$ and transmission eigenchannel $T_n$ are given in Supplemental Material\cite{suppsee}). From Fig. \ref{Fig1}(f), we can know that the disorder averaged transmission $\langle T \rangle$ is quantized over a wide range against disorder strength $U \in [0,3]t_1$ and its corresponding transmission fluctuation $\textrm{rms}(T)$ is zero presented in the insert figure. When the system length is increased from $L=100$ to $L=300$, the disorder averaged transmission $\langle T \rangle$ does not drop in the quantized region, which indicates the robustness of antichiral edge states. It is worth mentioning that the robustness of antichiral edge states depends on system's next-nearest hopping parameter $t_2$. The quantized transmission region can be enlarged up to $U=4.5$ when the magnitude of $t_2$ is doubled. Furthermore, the robustness of antichiral edge states against different corner shapes are presented in Supplemental Material\cite{suppsee}. In a nutshell, our proposed model to realize antichiral edge states is topologically protected and robust again random disorders.

\begin{figure}
\centering
\includegraphics[scale=0.40]{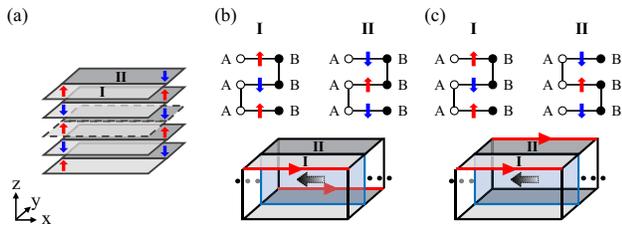}
\caption{(a) Schematic plot of the proposed three-dimensional model system stacking along $z$ direction, which consists of layered two-dimensional finite systems shown in Fig. \ref{Fig1}(a). The perpendicular red and blue arrows in the region I and II denote positive and negative magnetization in I and II regions, respectively. (b),(c) Schematic plot of antichiral hinge states denoted by the red arrows propagating along $x$ direction on the two opposite diagonal hinge cases, respectively. The corresponding hopping between two-dimensional layers are also shown. The counterpropagating states are denoted by the dark gray arrow.}
\label{Fig2}
\end{figure}

Next, let us consider how to achieve the antichiral hinge states in a three-dimensional system, which also propagate along the same direction. As shown in Fig. \ref{Fig2}(a), the model system is building by alternate stacking two-dimensional systems $H_{2D}$ with the Chern number $=\pm1$ as discussed in the previous section along $z$ direction. For the three-dimensional system, the nearest hopping term between adjacent layers is introduced\cite{tanaka2020theory}
\begin{eqnarray}
H_{z}
&=& \frac {t_{z}}{2}\sum _{i\in A,\alpha}(1-(-1)^{\alpha_A})c_{i\alpha}^\dagger c_{i\alpha+1}\nonumber \\
&+& \frac {t_{z}}{2}\sum _{i\in B,\alpha}(1+(-1)^{\alpha_B})c_{i\alpha}^\dagger c_{i\alpha+1} + H.c.,
\label{X3gc1}
\end{eqnarray}
where $\alpha_{A/B}$ represents $A/B$ site in $\alpha$th layer along the $z$ direction, $t_z$ is the nearest interlayer hopping strength. From Eq. (\ref{X3gc1}), we can know that the interlayer coupling terms give out either only A or B site connection between neighboring layers. The Hamiltonian for the overall three-dimensional system is given by $H = H_{2D} + H_{z}$.

Before investigating the electric properties of antichiral hinge states, we first discuss its topological behavior in the separate I/II region. Actually, the layered antiferromagnetic I/II system is a three-dimensional high-order topological insulator where chiral hinge states can emerge in the system. Note that the magnetization in each layer can be either $\uparrow$ or $\downarrow$ depending on the sign of phase $\phi$ in Eq. (\ref{Ham}). Moreover, the topological property of the bulk system can be characterized by the slab Chern number $C^z_{slab}$\cite{tanaka2020theory}. For instance, when the number of layers $N$ in the system is odd, the inversion symmetry is preserved [the inversion center is denoted as dashed line as shown in Fig. \ref{Fig2}(a)] and the slab Chern number is equal to $C^z_{slab}=\pm1$ depending on stacking order. On the other hand, when the number of layers $N$ in the system is even, the inversion symmetry is broken and the slab Chern number $C^z_{slab}=0$.
\begin{figure}
\centering
\includegraphics[scale=0.45]{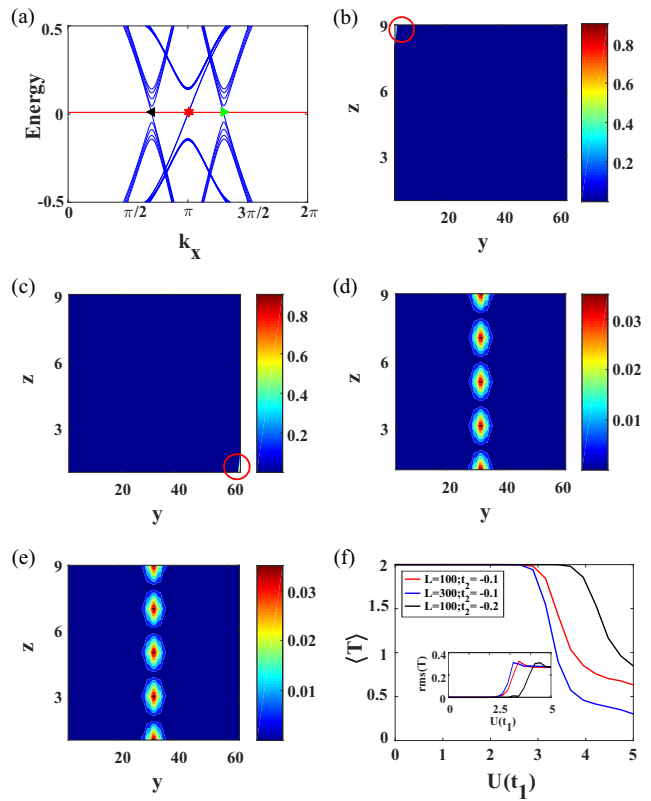}
\caption{(a) Band structure of the proposed three-dimensional antichiral hinge states system along $x$ direction shown in Fig. \ref{Fig2}(b). (b),(c) The absolute value of wave functions in $y-z$ plane for two antichiral hinge modes with $k_{x} = 3.16$ denoted as red up-pointing and down-pointing triangles in (a). (d),(e) The absolute value of wave functions in $y-z$ plane for two counterpropagating modes denoted as black left-pointing ($k_{x} = 2.19$) and green right-pointing ($k_{x} = 4.08$) triangles in (a). (f) Disorder averaged transmission coefficient $\langle T \rangle$ and its fluctuation $\textrm{rms}(T)$ versus disorder strength $U$ with two system lengths $L=100,300$. Each data point on the figure is averaged over $2000$ disorder configurations. Here, the parameters are $E = 0.01$, $N =9$, $W = 64$. $t_{1}=1$ is taken as the unit of energy, $t_{2} = -0.1$, $t_z$= 0.3.}
\label{Fig3}
\end{figure}
\begin{figure}
\centering
\includegraphics[scale=0.45]{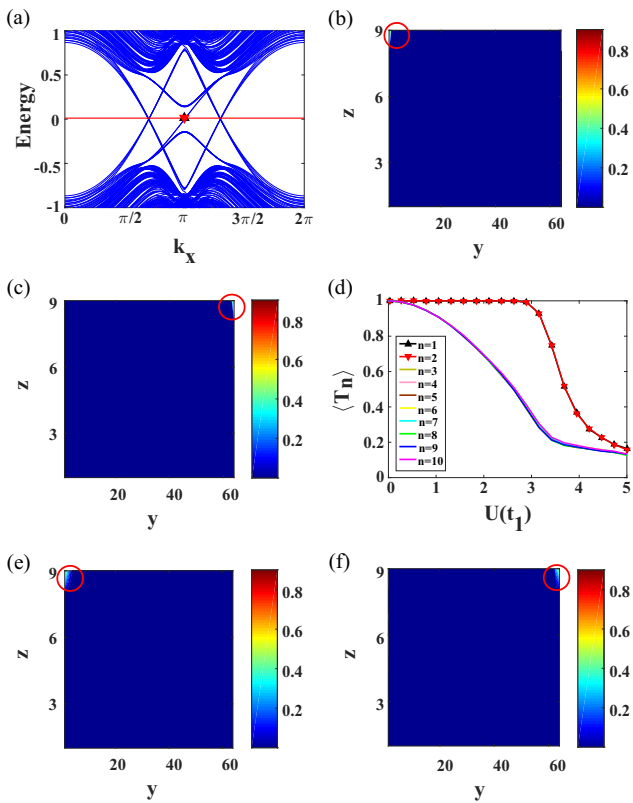}
\caption{(a) Band structure of the proposed three-dimensional antichiral hinge states system along $x$ direction shown in Fig. \ref{Fig2}(c). (b),(c) The absolute value of wave functions in $y-z$ plane for two antichiral hinge modes with $k_{x} = 3.16$ denoted as black up-pointing and red down-pointing triangles in (a). (d) The disorder averaged transmission eigenchannels $\langle T_n \rangle$ versus disorder strength $U$ with $L=100$ where $n=1,2,...,10$. Each data point on the figure is averaged over $2000$ disorder configurations. (e),(f) The absolute value of wave functions in $y-z$ plane when $x=100$ for two antichiral hinge modes ($n=1,2$) in a specific disorder configuration when $U=2$. Here, the parameters are $E = 0.01$, $N =9$, $W=64$. $t_{1}=1$ is taken as the unit of energy, $t_{2} = -0.1$, $t_z$= 0.3.}
\label{Fig4}
\end{figure}
In order to achieve the antichiral hinge states in the three-dimensional system, we particularly consider the system stacking with an odd number of layers. In the individual two-dimensional layer system, the phase $\phi_{I}$ and $\phi_{II}$ have same magnitude while have opposite sign with each other. For instance, from bottom layer to top layer, the phase $\phi_{I}$ in different layers are ${+\pi/2}$, ${-\pi/2}$,$\cdots$, while phase $\phi_{II}$ in different layers are ${-\pi/2}$, ${+\pi/2}$,$\cdots$, which gives out the corresponding positive or negative two-dimensional system's magnetization denoted by the red and blue arrows shown in Fig. \ref{Fig2}(a). As shown in Fig. \ref{Fig2}(b), we consider the model system is periodic along $x$ direction and is finite along $y$ and $z$ directions. The hopping parameters $\alpha_{A}$ and $\alpha_{B}$ are both equal to $\alpha$ in both regions as shown in Fig. \ref{Fig2}(b). Without loss of generality, the system parameters are chosen as layer number $N=9$ and width along $y$ direction $W=64$. The corresponding band structure is presented in Fig. \ref{Fig3}(a). Similar to the two-dimensional system, there are two degenerate states and two counterpropagating states with different momentum located in the band gap. By choosing the electron energy $E=0.01$, we present the absolute value of wave function distribution in $y-z$ plane in Figs. \ref{Fig3}(b)-\ref{Fig3}(e). It is clear that the two degenerate states are located in the upper left and lower right corners in $y-z$ plane as shown in Figs. \ref{Fig3}(b) and \ref{Fig3}(c). Since these two states have same momentum $k_x = 3.16$ and are propagating along positive $x$ direction, they are definite antichiral hinge states. Correspondingly, the other two counterpropagating states are mainly distributed in the central interface between region I and II presented in Figs. \ref{Fig3}(d) and \ref{Fig3}(e).
To test and verify the robustness of the antichiral hinge states, the random disorders are placed in the three-dimensional system with two length $L=100,300$ that is contacted with two semi-infinite leads. Note that the transmission coefficient is equal to two when there is no disorder, i.e., $U=0$. As the disorder strength is increasing up to 2.5$t_1$, the disorder averaged transmissions $\langle T \rangle$ for both systems are still quantized as two. Correspondingly, the transmission fluctuation $\textrm{rms}(T)$ is zero over this disorder strength region. Compared with short system ($L=100$), the quantized value is slightly decreased as the system length is tripled. This indicates that the proposed antichiral hinge states are robust against random disorders.

Lastly, we discuss the case of another antichiral hinge state as presented in Fig. \ref{Fig2}(c) where two hinge states located in the two hinges in the same $z$ plane. In this case, the hopping parameters $\alpha_{A}$ and $\alpha_{B}$ are equal to $\alpha$ in the region I while they are equal to $\alpha+1$ in the region II, which are shown in Fig. \ref{Fig2}(c). The corresponding band structure is given in Fig. \ref{Fig4}(a). Interestingly, we find that there is no band gap in the system. However, there are still two degenerate states propagating along the positive $x$ direction denoted as black up-pointing and red down-pointing triangles in Fig. \ref{Fig4}(a). By investigating the real-space distribution of the corresponding wave function with $E=0.01$ in the $y-z$ plane in Figs. \ref{Fig4}(b) and \ref{Fig4}(c), we easily know that these two states are antichiral hinge states in the same $z$ plane. In the meantime, other states at the same energy propagating along positive $x$ direction are bulk states as provided in the Supplemental Material\cite{suppsee}. Because the bulk states and antichiral hinge states are mixed, the disorder averaged transmission $\langle T_n \rangle$ for all channels versus disorder strength are studied in Fig. \ref{Fig4}(d). It is found that two transmission channels are quantized over a wide range of disorder strength while the transmission channels for other states quickly drop from one as the disorder strength increases, which are easily localized. To further present the robustness of antichiral hinge states, the corresponding absolute value of wave functions in $y-z$ plane when $x = 100, U=2$ for a specific disorder configuration are still localized in the corners shown in Figs. \ref{Fig4}(e) and \ref{Fig4}(f).

To summarize, we have proposed a theoretical model to realize topologically protected antichiral edge states in two dimensions and hinge states in three dimensions. The model is basically constructed by the Haldane model, which indicates that our system is topologically protected by choosing proper parameters. Their topological properties and robustness are investigated by calculating local Chern number and disordered transmission function. Since the antichiral edge states are recently achieved in the photonic system\cite{zhou2020observation} and electric system\cite{yang2020observation}, it is expected our proposed model can also be realized in the near future with the same experimental technique.

We gratefully acknowledge the support from National Key R\&D Program of China under Grant No. 2017YFA0304203, the National Natural Science Foundation of China (Grant No. 12074230), 1331KSC, Shanxi Province 100-Plan Talent Program. This research was partially conducted using the High Performance Computer of Shanxi University.

\bigskip

\noindent{$^{*)}$chenjun@sxu.edu.cn}\\
\noindent{$^{\dagger)}$zhanglei@sxu.edu.cn}

\bibliography{ref}

\end{document}